\documentclass[aps,pre,twocolumn,amsmath,amssymb,groupedaddress,showpacs,showkeys]{revtex4} 
\usepackage{graphicx}
\usepackage{subfigure}
\usepackage{bm}

\newtheorem{proposition}{Proposition}

\newcommand{\be}{\begin{equation}}
\newcommand{\ee}{\end{equation}}
\newcommand{\bea}{\begin{eqnarray}}
\newcommand{\eea}{\end{eqnarray}}
\newcommand{\bw}{\begin{widetext}}
\newcommand{\ew}{\end{widetext}}

\newcommand{\mm}{\mathrm}

\newcommand{\mc}{\mathcal}
\newcommand{\nn}{\nonumber\\}
\newcommand{\bi}{\begin{itemize}}
\newcommand{\ei}{\end{itemize}}
\newcommand{\etal}{\textit{et al }}

\begin{document}

  \title{No-passing Rule in the Ground State Evolution of the Random-Field Ising Model}
  \author{Yang Liu}
  \author{Karin A.\ Dahmen}
  \affiliation{University of Illinois at Urbana-Champaign, Department of
    Physics, 1110 West Green Street, Urbana, IL 61801.}

  \date{\today}

  \begin{abstract}
    We exactly prove the no-passing rule in the ground state evolution of the
    random-field Ising model (RFIM) with monotonically varying external
    field. In particular, we show that the application of the no-passing rule
    can speed up the calculation of the zero-temperature equilibrium $M(H)$
    curve dramatically. 
  \end{abstract}



  \maketitle

  \section{\label{sec:intro}Introduction}
  The no-passing rule was first introduced by Middleton in the study of
  sliding charge-density waves (CDW's)~\cite{Middleton-92}. The CDW problem
  belongs to the more general class of driven elastic manifolds in random
  media. If one defines a simple one-dimensional order parameter within the
  model, then a \emph{natural partial ordering} of the configurations can be
  defined. In the simple CDW model considered by Middleton, the CDW
  configuration $\{\varphi_i(t)\}$ describes the CDW distortions at $N$
  lattice sites indexed by $i$, with $\varphi_i(t)$ real phase variables and
  $t$ the time. The equation of motion for an overdamped CDW is governed by
  the Langevin dynamics  

  \be 
  \dot{\varphi_i} = \Delta^2 \varphi_i - V^\prime_i(\varphi_i) + f(t)
  \ee Here, the $\Delta^2 \varphi_i$ term represents the simple elastic
  interactions. $V^\prime_i(\varphi_i)$ is the pinning force at site $i$ due
  to the $2\pi$ periodic pinning potential $V(\varphi_i)$. And $f(t)$
  stands for the external driving force. Then one can define the natural
  partial ordering of two configurations: 
  $C^\mm{G} = \{\varphi^\mm{G}_i\} \ge C^\mm{L} = \{\varphi^\mm{L}_i\}$ if
  $\varphi^\mm{G}_i \ge \varphi^\mm{L}_i$ for each site $i$ of the
  system. The no-passing rule states that given a driving force $f$ if
  initially $C^\mm{G}(0) \ge C^\mm{L}(0)$, then $C^\mm{G}(t) \ge C^\mm{L}(t)$
  for all $t>0$, i.e. the ``greater'' ($C^\mm{G}$) is never passed by the
  ``lesser'' ($C^\mm{L}$). As stressed by Middleton, this rule relies crucially
  on the elastic potential being \emph{convex}. In other words, the elastic
  potential tends to decrease the separation of nearest-neighbor
  $\varphi$'s. More recently, Krauth \etal found a similar no-passing rule in
  the study of driven elastic strings in disordered
  media~\cite{Krauth-01,Krauth-06}. Obviously, this is the same general
  problem. Again, the rule is crucially dependent on the fact that the elastic
  potential is \emph{convex}. 

  The no-passing rule can be used to prove many useful properties, such as the
  asymptotic uniqueness of the sliding state for CDW's~\cite{Middleton-92} and
  the intriguing memory effects~\cite{Sethna-93}. In fact, just after its
  introduction by Middleton, the no-passing rule was used in the
  non-equilibrium zero-temperature random-field Ising model (RFIM) by Sethna
  \etal to prove the return point memory effect~\cite{Sethna-93}. The RFIM is
  obtained by adding a random field $h_i$ at each site of the Ising model  

  \be {\cal H} = - \!\sum_{{<}i,j{>}} J \, s_i s_j - \sum_i \, (H + h_i) \,
  s_i \label{eq:RFIM-Hamiltonian} \ee The distribution of $h_i$ values is
  usually taken to be Gaussian, with standard deviation $R$ and mean 0. $J$ is
  the nearest-neighbor ferromagnetic coupling strength and $H$ is the uniform
  external field. In this case, the natural partial ordering of two
  configurations can be defined similarly as in the CDW case. A difference is
  that $s_i=\pm 1$ while $\varphi_i$ is real. The no-passing rule states: Let
  a system $C^\mm{G}(t)$ be evolved under the fields $H^\mm{G}(t)$ and
  similarly $C^\mm{L}(t)$ evolved under $H^\mm{L}(t)$. Suppose the fields
  $H^\mm{G}(t)\ge H^\mm{L}(t)$ and the initial configurations satisfy
  $C^\mm{G}(0)\ge C^\mm{L}(0)$, then $C^\mm{G}(t)\ge C^\mm{L}(t)$ at all times
  later $t>0$, i.e. the partial ordering will be preserved by the
  dynamics. With a local metastable single-spin-flip dynamics, i.e. a spin
  flips when its effective local field \be h^{\rm eff}_i = J \sum_{j} s_j +
  h_i + H \label{eq:localField} \ee changes sign, the proof of the no-passing
  rule is straightforward~\cite{Sethna-93}. Even with a two-spin-flip
  dynamics, it has been shown by Vives \etal that the no-passing rule is still
  true at zero-temperature~\cite{Vives-05}. Note that for the magnetization
  process, the no-passing rule is equivalent to the fact that the flipped
  spins can never flip back as $H$ is swept monotonically.  Again, this rule
  is not unconditionally true. It relies crucially on the nearest-neighbor
  interaction being \emph{ferromagnetic} ($J>0$). Just like the convex elastic
  potential, the ferromagnetic interaction also tends to decrease the
  separation of nearest-neighbor degrees of freedom, i.e. it tends to align
  the spins.

  Recently, in the study of the equilibrium zero-temperature RFIM, Vives \etal
  conjectured that when the external field $H$ is swept from $-\infty$ to
  $\infty$, flipped spins cannot flip back in the equilibrium $M(H)$
  curve~\cite{Vives-02}. In other words, the no-passing rule is valid even for
  the zero-temperature equilibrium dynamics, i.e. the evolution of the ground
  state (GS). Vives \etal further conjectured that this rule can be used to speed up the calculation
  of the equilibrium $M(H)$ curve since flipped spins at a lower field can
  be removed from the GS calculation for all higher fields. Unfortunately,
  this simple but powerful rule has not been proven so far for the equilibrium
  RFIM. This is the main motivation of our work.

  This paper is organized as follows. In Sec.II, we give a short introduction
  to the calculation of the equilibrium $M(H)$ curve of the zero-temperature
  RFIM. In Sec.III, we work out some basic steps for the proof of the
  equilibrium no-passing rule. In Sec.IV, we present the proof. In Sec.V, we
  show the direct application of this rule to the calculation of the
  equilibrium $M(H)$ curve. Finally, in Sec.VI we discuss its validity in
  other systems.

  \section{\label{sec:algorithm} the equilibrium $M(H)$ curve}

  To calculate the equilibrium $M(H)$ curve of the zero-temperature RFIM, we
  first need to calculate the exact GS in the RFIM at an arbitrary applied
  external field $H$. This is the basic step of calculating the 
  equilibrium $M(H)$ curve, i.e. the GS evolution for varying $H$. Fortunately,
  there is a well-known mapping of the RFIM GS problem to a min-cut/max-flow
  problem in combinatorial optimization. The mapping and the so-called
  push-relabel algorithm for the min-cut/max-flow problem has been well
  described in the literatures~\cite{Hartmann-02,Cherkassky-97}. For RFIM, the
  run time of the push-relabel algorithm scales as $\mc{O}(N^{4/3})$ with $N$
  the system size~\cite{Hartmann-98,Middleton-01}.

\begin{figure*}[t]
    \includegraphics[width=6.0in]{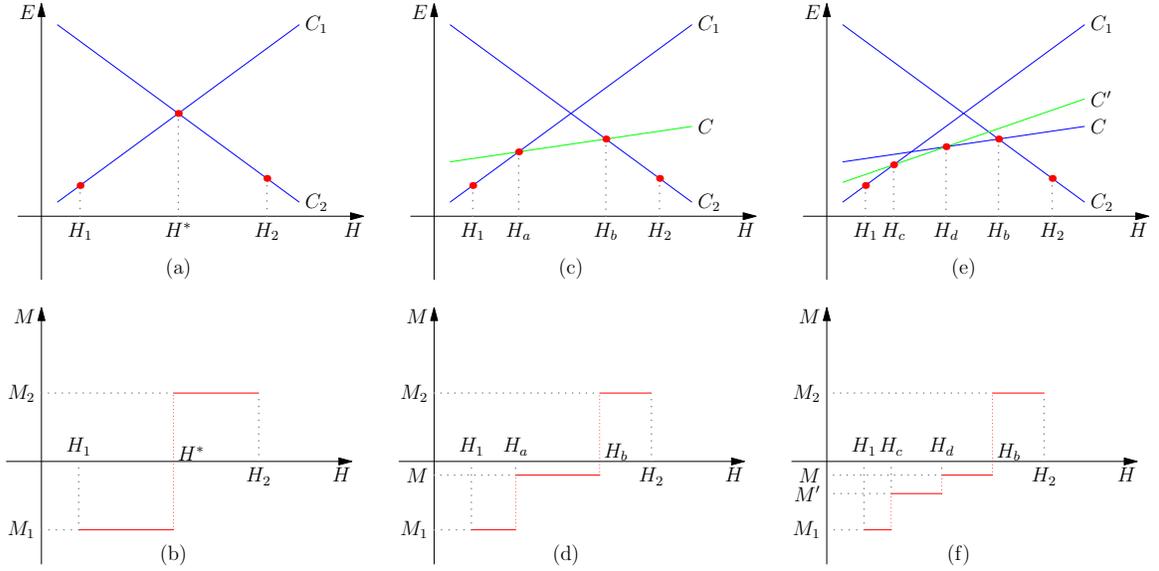}
    \caption{\label{fig:algorithm} An illustration of the algorithm to
      calculate the equilibrium $M$-$H$ curve. Calculate the energies $E_1$
      and $E_2$ of the two simplest states $C_1=\{s_i=-1\}$ and
      $C_2=\{s_i=+1\}$, respectively, as a function of $H$. According to
      Proposition.~\ref{pro:1}, $C_1$ (or $C_2$) would be the ground state for
      $H<-h_\mm{max}$ (or $H>-h_\mm{min}$). Calculate the crossing field
      $H^*(C_1,C_2)$ where $E_1=E_2$. Check whether there is a GS at $H^*$
      which is different from $C_1$ and $C_2$. If no, the algorithm ends. If
      yes, denote the GS as $C$, calculate the crossing field $H^*(C,C_1)$ and
      $H^*(C,C_2)$. At the new crossing fields, check whether there is a GS
      which is different from the two intersected states. The algorithm will
      not end until all the crossing fields have been checked. An example of
      the calculated equilibrium $M$-$H$ curve is shown in Fig.~\ref{fig:MH}.}        
  \end{figure*}

\begin{figure}[t]
    \includegraphics[width=0.48\textwidth]{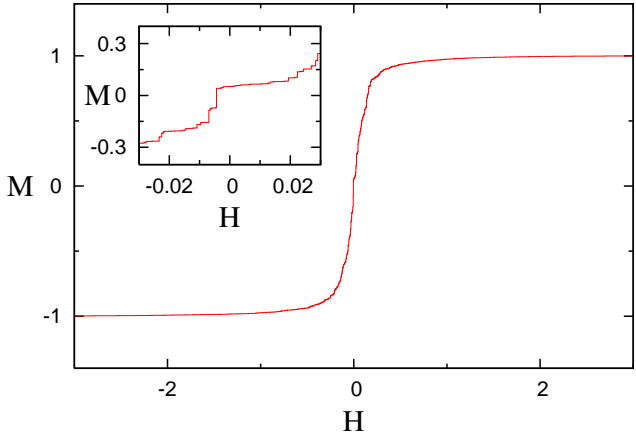}
    \caption{\label{fig:MH} The equilibrium $M$-$H$ curve (the ground state
      evolution) for the Gaussian RFIM with $D=3$, $L=32$ and $R=2.837$. Here
      $R$ is the standard deviation of the Gaussian random field
      distribution. The inset shows a detail of the $M(H)$ curve near $H=0$,
      where magnetization jumps are clearly seen. }
  \end{figure}

  The equilibrium $M(H)$ curve can be simulated with the method reported 
  in Ref.~\onlinecite{Hartmann-98,Vives-00}. It is essentially based on the
  fact that the GS energy $E({S_i},H)$ is convex up in $H$, which allows for
  estimates of the fields $H$ where the magnetization jumps (called
  ``avalanches'' occur). This algorithm finds steps by narrowing down ranges
  where the magnetization jumps with an efficient linear interpolation scheme.      
  An illustration of the algorithm is shown in Fig.~\ref{fig:algorithm}. An
  example of the calculated equilibrium $M(H)$ curve is shown in
  Fig~\ref{fig:MH}. The details have been explained extensively in
  Ref.~\onlinecite{Vives-00}. Here, we just introduce some important
  propositions.

  In the $E-H$ diagram, for each state $\{s_i\}$, the energy $E$ is represented by
  a straight line with slope $-M\equiv -\sum_i s_i$ since 
  \be E(\{s_i\},H)=E_0(\{s_i\})-HM \ee with 
  \be E_0(\{s_i\})= - \!\sum_{{<}i,j{>}} J \, s_i s_j - \sum_i \,  h_i \, s_i
  \ee the energy axis intercept, i.e. the total energy of the configuration
  when $H=0$. Consider a D-dimensional hypercubic lattice of size $N=L^D$. Let
  $h_\mm{max}$ ($h_\mm{min}$) be the maximum (minimum) values of $h_i$ for a
  certain realization of the random fields. (Usually, random fields are chosen
  from a Gaussian distribution with mean 0 and standard deviation $R$. $R$ is
  often called the disorder parameter.) Most of the following simple
  propositions have been proven in Ref.~\onlinecite{Vives-00}.   

  \begin{proposition}\label{pro:1}
    For $H<-h_\mm{max}$ $(H>-h_\mm{min})$, the ground state is $\{s_i=-1\}$
    $(\{s_i=+1\})$.  
  \end{proposition}

  \begin{proposition}\label{pro:Monotonical}
    Let the spin configuration $C_1$ $(C_2)$ be the ground state for $H=H_1$
    $(H=H_2)$. They have magnetization $M_1$ and $M_2$, respectively. If $C_1
    \ne C_2$ and $H_2 >H_1$, then $M_2 > M_1$.  
  \end{proposition}

  Thus, when sweeping the external field from $H=-\infty$ to $H=\infty$, the
  magnetization $M$ will increase monotonically. A corollary of this
  proposition is that: In the $E-H$ diagram if the slopes of the lines
  corresponding to the ground states $C_1$ and $C_2$ are different, i.e. $M_1
  \ne M_2$, and without loss of generality we can assume $M_1 < M_2$, then the 
  lines intersect at a field $H^{*}$ such that $H_1 < H^{*} < H_2$. This field
  $H^{*}(C_1,C_2)$ is defined as the \emph{crossing field} between $C_1$ and
  $C_2$. According to the definition, one has $E_0(C_1)-H^*M_1 =
  E_0(C_2)-H^*M_2$, so    
  \be H^{*}(C_1,C_2) = \frac{E_0 (C_2) - E_0  (C_1)}{M_2 - M_1} \ee For
  example, we can calculate the crossing field between the two simplest ground
  states: $C_1=\{s_i=-1\}$ with $M_1=-N$ and $C_2=\{s_i=+1\}$ with $M_2=N$. We
  have $H^{*} =-1/N \sum_i h_i = -\bar{h}_i$. For a zero-mean distribution of
  the random fields, we should have $H^{*}=0$.

  \begin{proposition}\label{proposition:key}
    Let the spin configuration $C_1$ $(C_2)$ be the ground state for $H=H_1$
    $(H=H_2)$.  $C_1 \ne C_2$, $H_2 >H_1$ and the crossing field between $C_1$
    and $C_2$ is $H^{*}$. If there is no configuration $C$ such that $E(C,H^*)
    < E(C_1,H^* )=E(C_2,H^* )$ then: (i) $C_1$ is the ground state at least
    for the field range $[H_1,H^*)$ and (ii) $C_2$ is the ground state at
    least for the field range $(H^*,H_2]$. 
  \end{proposition}

  This is the most important proposition. Its power comes from the fact that
  it can be applied iteratively. And finally we get the $M(H)$ curve with all
  the ground states. See Fig.~\ref{fig:algorithm}.

  \begin{proposition}\label{pro:single}
    If the ground state is non-degenerate, then there can not be more than one
    avalanche connecting two nearest ground states in the $E-H$ diagram.   
  \end{proposition}

  The proof of this interesting proposition is shown in the Appendix. This
  proposition says that if the GS is non-degenerate, when we increase the
  external field $H$ adiabatically slowly, we can trigger just one avalanche
  at a time.

  \section{\label{sec:lemma}Preparations of the Proof}

  In this section, we will work out the total energy change of the spin
  configuration due to multiple spin flips and external field varying. The
  spin configuration is not necessarily the ground state. 

  First, let's consider the simplest case of a single spin flip. Suppose only
  one spin ($s_i$) flips during the evolution of configuration $C$ at $H$
  to configuration $C'$ at $H'$, with $\Delta H= H' - H$ and $\Delta M= M' -
  M$. Define $n_i$ (or $n'_i$) to be the number of the $i$-th spin's nearest
  neighbors that point in the same direction as the spin in configuration $C$ (or
  $C'$). We call these spins the same-direction nearest neighbors(SDNN) of the
  $i$-th spin. Note that $n_i = 0,1,2,...Z$ with $Z=2D$ the coordination
  number of the $D$-dimensional hypercubic lattice.

  It is easy to get the bond energy change $4J(n_i-D)$. And the total energy
  change due to the single spin-flip and the varying external field is given
  by     
  
  \be f_{i}(H, \Delta H) =  f_i(H) - \Delta HM'\ee   Here we have defined 
  \be f_i(H) \equiv f_{i}(H,0) = 4J(n_i - D) - (h_i+H) \Delta s_i  \ee
  which is the energy change due to spin $i$ flipping for the configuration $C$ just at
  the field $H$, i.e. $\Delta s_i=\pm 2$ with $\Delta H=0$. It is easy to
  check that \be f_{i,\pm}(H) = f_{i,\pm}(0) \pm 2H = \pm 2 h^{\rm  eff}_i
  \label{eq:fiH} \ee with `$\pm$' represents $s_i=\pm1$ and $\Delta s_i = \mp
  2$ accordingly.

  Second, we consider two spin flips. Suppose two different spins ($s_i$
  and $s_j$) flip during the evolution of configuration $C$ at $H$ to configuration
  $C'$ at $H'$. There are two subcases. 

  (1) $s_i$ and $s_j$ are not next to each other. The energy change is 

      \be f_{i,j}(H,\Delta H) = f_i(H) + f_j(H) -\Delta H M' \ee 

  (2) $s_i$ and $s_j$ are next to each other. The energy change is 
      
      \be f_{\langle i,j \rangle }(H,\Delta H)= f_i(H) + f_j(H) - 4J (s_i\cdot
      s_j) - \Delta H M' \ee Note that the term $-4J (s_i\cdot s_j)$ is just
      due to the fact that the energy of the $i-j$ bond will not change during
      the flip.

  \begin{figure*}[t]
    \includegraphics[width=6in]{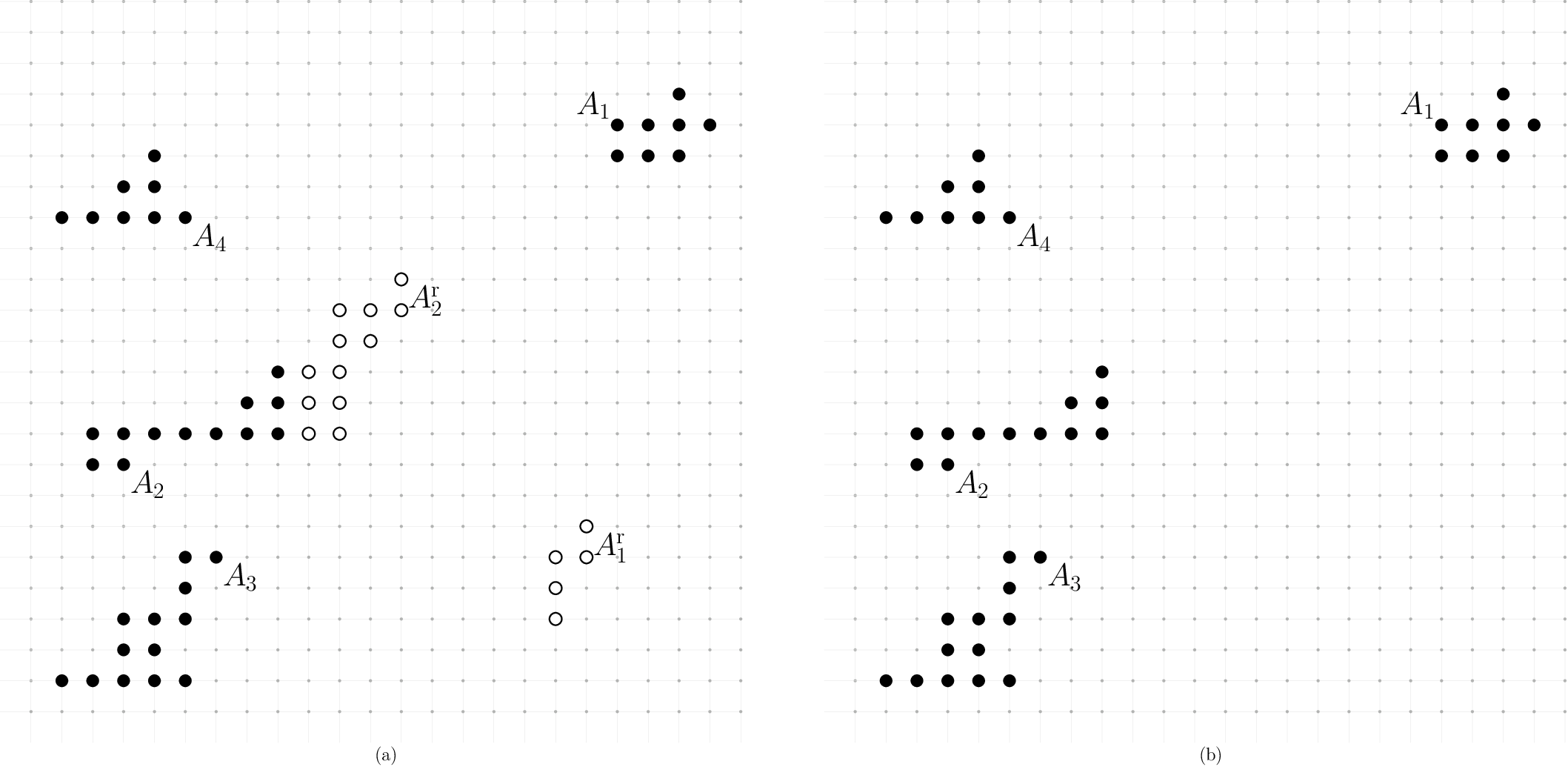}
    \caption{\label{fig:manyavalanches} States evolved from $C_1$, with only
    the change of the spin configuration, i.e. avalanches and reverse
    avalanches are explicitly shown. (a) State $C_2$: evolved from state $C_1$
    with both avalanches and reverse avalanches. (Black dot) spins flip UP,
    forming avalanches ($A_1$, $A_2$, $A_3$ and $A_4$). (White dot) spins flip
    DOWN (reverse flip), forming reverse avalanches ($A^\mm{r}_1$,
    $A^\mm{r}_2$). Note that there are three interacting bonds between
    avalanche $A_2$ and reverse avalanche $A^\mm{r}_2$. (b) State $\tilde{C}$
    evolved from state $C_1$ without reverse avalanches.}   
  \end{figure*}

  Finally, let's consider the general case. See
  Fig.~\ref{fig:manyavalanches}(a). There are many spin flips during the 
  evolution of configuration $C$ at field $H$ to $C'$ at field $H'$. It is easy to
  check that the total energy change is given by

  \bea \Delta E(H,\Delta H) &=& \big[ f_i(H) + f_j(H) + \cdots \big]\nn
                            & & - 4J \ (s_i\cdot s_j + \cdots) \nn
                            & & -  \Delta H (M+\Delta s_i + \Delta s_j + \cdots)
  \label{eq:generalDeltaE}\eea
  On the RHS, the first term includes all the flipping spins. The second term
  includes all the nearest-neighbor interactions among those flipping
  spins. The last term is due to the varying external field. In particular, if
  all the flipping spins flip at the same $H$ and they are connected to each
  other and have the same spin value $-1$ (or $+1$) before the flip, then this
  collective spin flip is called \emph{ an avalanche} (or \emph{a reverse
  avalanche}). 

  Denote the energy change due to an avalanche $A_\alpha$ as
  $f_{A_\alpha}(H,\Delta H)$, we have

  \bea f_{A_\alpha}(H,\Delta H) & = & \big[f_{i}(H) + f_{j}(H) + \cdots \big] - 4J N_{\mm{b}}(A_\alpha) \nn
                                &   &    -  \Delta H (M+2 S_\alpha) \nn  
                           & \equiv & f_{A_\alpha}(H) - \Delta H (M+2 S_\alpha) 
  \eea with $N_\mm{b}(A_\alpha)$ defined as the number of interacting bonds in
  $A_\alpha$, $S_\alpha$ the size of the avalanche and $f_{A_\alpha}(H)$ the
  energy change due to the avalanche when $\Delta H=0$. Similarly, for the
  reverse avalanche, we have  

  \bea f_{A_\beta^\mm{r}}(H,\Delta H) & = & \big[f_{i}(H) + f_{j}(H) + \cdots \big] - 4J N_\mm{b}(A_\beta^\mm{r})\nn
                                  &   &  -  \Delta H (M-2 S_\beta^\mm{r}) \nn 
                                  & \equiv & f_{A_\beta^\mm{r}}(H) - \Delta H
                                  (M- 2 S_\beta^\mm{r}) \eea 
  Due to Eq.~\ref{eq:fiH}, we have  

  \bea f_{A_\alpha}(H) &=& f_{A_\alpha}(0) - 2 S_\alpha H  \\
       f_{A_\beta^\mm{r}}(H) &=& f_{A_\beta^\mm{r}}(0) + 2 S_\beta^\mm{r} H
  \eea

  Now we can rewrite Eq.~\ref{eq:generalDeltaE} in terms of $f_{A_\alpha}$ and
  $f_{A{^\mm{r}_\beta}}$. The total energy change due to avalanches and
  reverse avalanches is given by     

  \bea \Delta E(H,\Delta H) 
  &=&  F_{A}(H) + F_{A^\mm{r}}(H)+ \ 4J  N_\mm{b}(A,A^\mm{r}) \nn
  & & - \ \Delta H (M+ 2 S_{A}-2S_{A^\mm{r}} ) 
  \label{eq:multipleavalanche} \eea 

  with notations $ F_{A}(H) \equiv \sum_{\alpha} f_{A_\alpha}(H)$, 
  $ F_{A^\mm{r}}(H) \equiv \sum_{\beta} f_{A^\mm{r}_\beta}(H)$,
  $ S_{A} \equiv \sum_{\alpha} S_\alpha$,
  $ S_{A^\mm{r}} \equiv \sum_{\beta} S^\mm{r}_\beta$. Here
  $N_\mm{b}(A,A^\mm{r})$ denotes the number of interacting bonds between
  avalanches and reverse avalanches. For example,
  in Fig.~\ref{fig:manyavalanches}(a),  $N_\mm{b}(A,A^\mm{r})=3$.

  \section{\label{sec:proof}Proof of the No-passing rule}
  Now we are ready for the proof of the no-passing rule. Let the spin
  configuration $C_1$ $(C_2)$ be the ground state for $H=H_1$ $(H=H_2)$.  $H_2
  >H_1$. Suppose $C_1$ and $C_2$ are connected with multiple avalanches: $A_1,
  A_2 \cdots A_n$ with sizes $S_1, S_2 \cdots S_n$ and reverse avalanches
  $A^\mm{r}_1, A^\mm{r}_2 \cdots A^\mm{r}_m$, with sizes
  $S^\mm{r}_1,S^\mm{r}_2, \cdots, S^\mm{r}_m$ respectively. To compensate
  these reverse avalanches (so as to make sure $M$ is monotonically
  increasing, see Proposition.~\ref{pro:Monotonical}),
  we must have $ S_A = \sum_{\alpha=1}^n S_\alpha > S_{A^\mm{r}} =
  \sum_{\beta=1}^m S^\mm{r}_\beta $. See Fig.~\ref{fig:manyavalanches}(a). 

  The idea is that if $C_2$ is the GS at field $H_2$, then it should have lower
  energy than any other spin configuration at $H_2$. But we will prove this
  is NOT true. Just consider another spin configuration $\tilde{C}$. The only
  difference between $C_2$ and $\tilde{C}$ is that $\tilde{C}$ is evolved
  from $C_1$ without any reverse avalanches. See
  Fig.~\ref{fig:manyavalanches}(b). We now try to prove that $E(\tilde{C},H_2)
  < E(C_2,H_2)$, so $C_2$ can NOT be the GS at $H_2$. But this is equivalent
  to proving  that $\Delta \tilde{E} < \Delta E$. Here, 
  
  \bea \Delta E & \equiv & E(C_2,H_2) - E(C_1,H_1) \nn
  & = &  F_{A^\mm{r}}(H_1) + F_{A}(H_1) + \ 4J \ N_\mm{b}(A,A^\mm{r}) \nn
  &   &  - \ \Delta H (M-2 S_{A^\mm{r}} + 2 S_{A}) 
  .  \eea On the other hand,   

  \bea \Delta \tilde{E} &\equiv& E(\tilde{C},H_2) - E(C_1,H_1) \nn
                        &=& F_{A}(H_1) - \Delta H (M+2 S_{A})  \eea Therefore, 
  
  \be \Delta E - \Delta \tilde{E} = F_{A^\mm{r}}(H_1) + 4J 
  N_\mm{b}(A,A^\mm{r}) + 2 S_{A^\mm{r}}  \Delta H > 0 \label{eq:methodI}\ee
  Here we have used the fact that $C_1$ is the ground state for $H=H_1$ such
  that any kinds of spin flip will increase the energy: $f_{A^\mm{r}_{\beta}}(H_1)
  >0  \Rightarrow F_{A^\mm{r}}(H_1) > 0$.  Also, for the ferromagnetic RFIM,
  $J>0$. Since each term is positive, so the sum is positive, i.e.  $\Delta E
  > \Delta \tilde{E}$ or $E > \tilde{E}$. Actually, for any state $C_2$ which
  evolved from $C_1$ with reverse avalanches, we can find a corresponding
  state $\tilde{C}$ which evolved from $C_1$ without any reverse avalanches
  that has lower energy than $C_2$ at field $H_2$. So reverse spin flips are
  impossible for ground state evolution when increasing external
  field. Generally, flipped spins can never flip back when we sweep the
  external field monotonically.


  \section{\label{sec:disc} Application}

\begin{figure}[t]
    \includegraphics[width=0.5\textwidth]{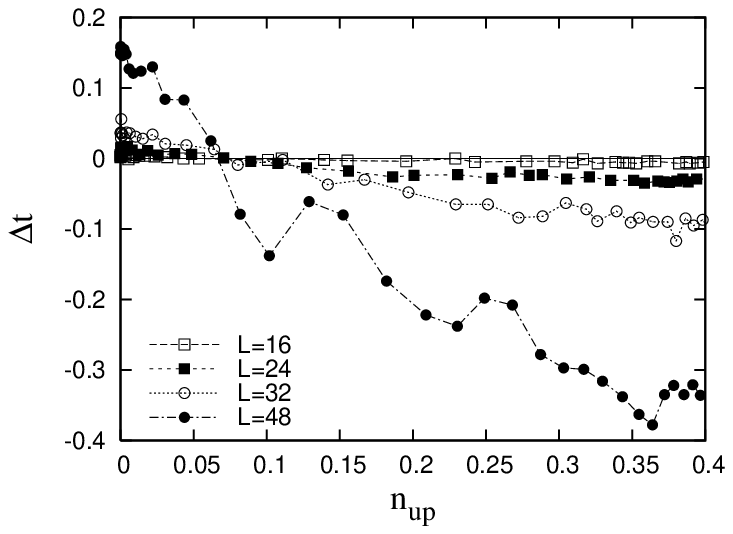}
    \includegraphics[width=0.5\textwidth]{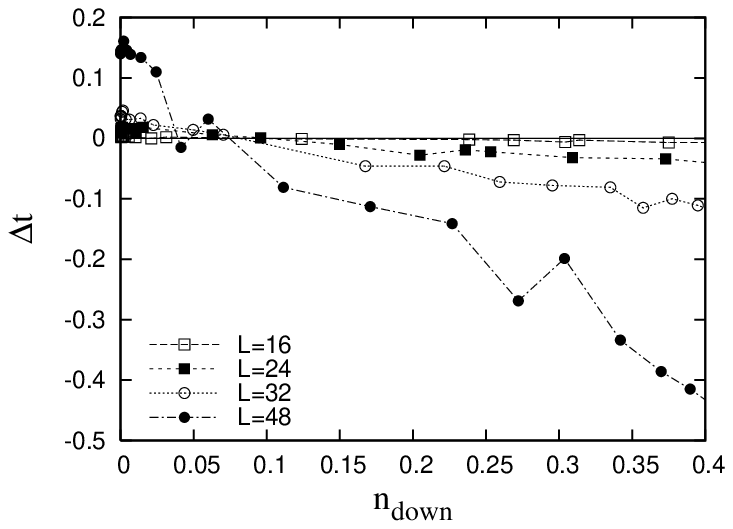}
    \caption{\label{fig:time} Running time difference ($\Delta t$) between
  methods with and without using the earlier solution, i.e the GS $C_1$ at
  field $H_1$, to calculate the GS $C_2$ at field $H_2$. The time difference
  $\Delta t$ (given in seconds) is plotted against the UP-spin (or DOWN-spin)
  density of the GS $C_1$. $\Delta t <0 $ means using the earlier solution
  will save the running time. Calculations are done for 3D Gaussian RFIM (with
  disorder parameter $R=2.27$) for different system sizes.  (Top) $H_2 >
  H_1$. UP spins in $C_1$ at field $H_1$ will stay UP at field $H_2$. (Bottom)
  $H_2 < H_1$. DOWN spins in $C_1$ at field $H_1$ will stay DOWN at field
  $H_2$.} 
  \end{figure}

  The straightforward application of the no-passing rule is very useful to accelerate the
  calculation of the ground states when varying the external field. Suppose the GS
  $C_1$ at field $H_1$ has already been obtained, and we want to calculate the GS $C_2$ at
  field $H_2$ with $H_2 > H_1$. According to the no-passing rule, the UP
  spins in $C_1$ will stay UP in $C_2$, i.e. those spins are frozen, so we
  needn't consider them in the ground state analysis. We just need to consider
  the DOWN spins in $C_1$. The only cost is that we have to deal with the
  frozen UP spins as complicated fixed boundary conditions for the DOWN
  spins. At first sight, one may think that only when the density of the
  frozen spins is big enough can we make the GS calculation faster. But how
  big is enough? To optimize our calculation, we consider the running time
  difference ($\Delta t$) between the two methods: (I) without using the earlier
  solution $C_1$; (II) using the earlier solution $C_1$. For both methods, ground
  states are found by using the push-relabel algorithm. The numerical
  experiments are conducted on a desktop with 2.80 GHz CPU and 2GB Memory. 
  And we tune  the UP-spin density $n_\mm{up}$ (Down-spin density
  $n_\mm{down}$) by varying $H_1$. The result is shown in
  Fig.~\ref{fig:time}. It is found that for $H_2 > H_1$, as long as $n_\mm{up}
  \gtrsim 0.07$ in GS $C_1$, method II will be faster than I. Symmetrically,
  for $H_2 < H_1$, as long as $n_\mm{down} \gtrsim 0.07$ in GS $C_1$, method
  II will be faster than I.  This suggests it is not necessary to have an
  extremely large portion of frozen spins to use the earlier
  solution. Freezing a tiny part of spins will accelerate the GS calculation
  already. Furthermore, for larger and larger density of the frozen spins, using
  the earlier solution will save more and more running time. (Keep in mind that
  for RFIM, the running time of the push-relabel algorithm scales as
  $\mc{O}(N^{4/3})$.) Consequently, the calculation of
  the whole $M(H)$ curve will be sped up dramatically.

  \section{\label{sec:disc} Discussions}

  Throughout our proof of the no-passing rule, we don't assume that the ground
  state is unique. In other words, the no-passing rule is correct even when
  the ground state is degenerate. For example, this happens for the RFIM when
  the random fields are chosen from a bimodal distribution~\cite{Hartmann-99}.  

  In the proof we explicitly use the fact that the nearest neighbor
  interaction should be \emph{ferromagnetic} ($J>0$). This means any
  antiferromagnetic interactions will destroy the no-passing rule. Thus, for
  other random magnet models, if $J_{ij}$ could be negative, such as the
  random-bond Ising model (RBIM) with negative $J_{ij}$ or the spin glasses,
  the rule will be violated.     

  Finally, we conjecture that for elastic manifolds in random media, there
  could be a similar equilibrium no-passing rule at zero temperature, provided
  that the elastic potential is convex and partial ordering of the
  configurations can be clearly defined.

  \section{Acknowledgments}

  We thank James P. Sethna, A. Alan Middleton and Werner Krauth for valuable
  discussions. We acknowledge the support of NSF Grant No. DMR 03-14279 and
  NSF Grant No. DMR 03-25939 ITR (Materials Computation Center). This work was
  conducted on the Beowolf cluster of the Materials Computation Center at
  UIUC.

  \appendix
  \section{\label{app:proof} A single avalanche connects two nearest ground states}

  Here, we show the proof of Proposition.~\ref{pro:single}.

  \textbf{Proof}: Suppose when the field is increased from $H_1$ to $H_2$, the GS $C_1$
  evolves to the nearest GS $C_2$ with two avalanches ($A_1$ and $A_2$ with
  size $S_1\ge1$ and $S_2\ge1$, respectively).

  The crossing field is given by 

  \bea H^{*}(C_1,C_2) &=& \frac{E_0 (C_2) - E_0 (C_1)}{M_2 - M_1} \nn
                      &=& \frac{f_{A_{1}}(0) + f_{A_{2}}(0)}{2(S_1+S_2)} 
  \label{eq:Hcross2}\eea

  The last line is due to Eq.\ref{eq:multipleavalanche}. We can choose a trial
  state $C$ which is evolved from $C_1$ with only avalanche $A_1$ occurring.  We
  want the following relation to hold

  \be E(C,H^* ) < E(C_1,H^* )=E(C_2,H^* ) \ee which is equivalent to 

  \bea E(C,H^* ) - E(C_1,H^* ) &=&  f_{A_{1}}(H^*) \nn
                               &=&  f_{A_{1}}(0) -2 S_1 H^* \nn
                               &<&  0 \eea \label{}

  Plugging Eq.~\ref{eq:Hcross2} in it, we just need to prove  

  \be S_2 f_{A_1}(0) < S_1  f_{A_2}(0) .\ee  

  If we do have $S_2 f_{A_1}(0) < S_1 f_{A_2}(0)$, then we choose the trial
  state $C$, which has lower energy than $C_1$ and $C_2$ at the field $H^*$;
  if $S_2 f_{A_1}(0) > S_1 f_{A_2}(0)$, then we can choose another trial state
  $C'$, which is evolved from $C_1$ through only avalanche $A_2$ and has lower
  energy than both $C_1$ and $C_2$ at $H^*$. In both cases, we have shown that
  $C_2$ can not be the nearest GS at $H_2$ for the GS $C_1$ at $H_1$, if $C_1$
  evolves to $C_2$ with two avalanches. If $S_2 f_{A_1}(0) = S_1
  f_{A_2}(0)$, it is easy to show that
  $E(C_1,H^*)=E(C,H^*)=E(C',H^*)=E(C_2,H^*)$, then there will be degenerate
  ground states at $H^*$, which is in contradiction to the hypothesis of
  Proposition.~\ref{pro:single}.  \\
  Q.E.D

  \bibliography{Nopassing}

\begin{thebibliography}{12}
\expandafter\ifx\csname natexlab\endcsname\relax\def\natexlab#1{#1}\fi
\expandafter\ifx\csname bibnamefont\endcsname\relax
  \def\bibnamefont#1{#1}\fi
\expandafter\ifx\csname bibfnamefont\endcsname\relax
  \def\bibfnamefont#1{#1}\fi
\expandafter\ifx\csname citenamefont\endcsname\relax
  \def\citenamefont#1{#1}\fi
\expandafter\ifx\csname url\endcsname\relax
  \def\url#1{\texttt{#1}}\fi
\expandafter\ifx\csname urlprefix\endcsname\relax\def\urlprefix{URL }\fi
\providecommand{\bibinfo}[2]{#2}
\providecommand{\eprint}[2][]{\url{#2}}

\bibitem[{\citenamefont{Middleton}(1992)}]{Middleton-92}
\bibinfo{author}{\bibfnamefont{A.~A.} \bibnamefont{Middleton}},
  \bibinfo{journal}{Phys. Rev. Lett.} \textbf{\bibinfo{volume}{68}},
  \bibinfo{pages}{670} (\bibinfo{year}{1992}).

\bibitem[{\citenamefont{Rosso and Krauth}(2001)}]{Krauth-01}
\bibinfo{author}{\bibfnamefont{A.}~\bibnamefont{Rosso}} \bibnamefont{and}
  \bibinfo{author}{\bibfnamefont{W.}~\bibnamefont{Krauth}},
  \bibinfo{journal}{Phys. Rev. B} \textbf{\bibinfo{volume}{65}},
  \bibinfo{pages}{012202} (\bibinfo{year}{2001}).

\bibitem[{\citenamefont{Kolton et~al.}(2006)\citenamefont{Kolton, Rosso,
  Giamarchi, and Krauth}}]{Krauth-06}
\bibinfo{author}{\bibfnamefont{A.~B.} \bibnamefont{Kolton}},
  \bibinfo{author}{\bibfnamefont{A.}~\bibnamefont{Rosso}},
  \bibinfo{author}{\bibfnamefont{T.}~\bibnamefont{Giamarchi}},
  \bibnamefont{and} \bibinfo{author}{\bibfnamefont{W.}~\bibnamefont{Krauth}},
  \bibinfo{journal}{Phys. Rev. Lett.} \textbf{\bibinfo{volume}{97}},
  \bibinfo{eid}{057001} (pages~\bibinfo{numpages}{4}) (\bibinfo{year}{2006}).

\bibitem[{\citenamefont{Sethna et~al.}(1993)\citenamefont{Sethna, Dahmen,
  Kartha, Krumhansl, Roberts, and Shore}}]{Sethna-93}
\bibinfo{author}{\bibfnamefont{J.~P.} \bibnamefont{Sethna}},
  \bibinfo{author}{\bibfnamefont{K.}~\bibnamefont{Dahmen}},
  \bibinfo{author}{\bibfnamefont{S.}~\bibnamefont{Kartha}},
  \bibinfo{author}{\bibfnamefont{J.~A.} \bibnamefont{Krumhansl}},
  \bibinfo{author}{\bibfnamefont{B.~W.} \bibnamefont{Roberts}},
  \bibnamefont{and} \bibinfo{author}{\bibfnamefont{J.~D.} \bibnamefont{Shore}},
  \bibinfo{journal}{Phys. Rev. Lett.} \textbf{\bibinfo{volume}{70}},
  \bibinfo{pages}{3347} (\bibinfo{year}{1993}).

\bibitem[{\citenamefont{Vives et~al.}(2005)\citenamefont{Vives, Rosinberg, and
  Tarjus}}]{Vives-05}
\bibinfo{author}{\bibfnamefont{E.}~\bibnamefont{Vives}},
  \bibinfo{author}{\bibfnamefont{M.~L.} \bibnamefont{Rosinberg}},
  \bibnamefont{and} \bibinfo{author}{\bibfnamefont{G.}~\bibnamefont{Tarjus}},
  \bibinfo{journal}{Phys. Rev. B} \textbf{\bibinfo{volume}{71}},
  \bibinfo{pages}{134424} (\bibinfo{year}{2005}).

\bibitem[{\citenamefont{Frontera and Vives}(2002)}]{Vives-02}
\bibinfo{author}{\bibfnamefont{C.}~\bibnamefont{Frontera}} \bibnamefont{and}
  \bibinfo{author}{\bibfnamefont{E.}~\bibnamefont{Vives}},
  \bibinfo{journal}{Comp.\ Phys.\ Comm.} \textbf{\bibinfo{volume}{147}},
  \bibinfo{pages}{455} (\bibinfo{year}{2002}).

\bibitem[{\citenamefont{Hartmann and Rieger}(2002)}]{Hartmann-02}
\bibinfo{author}{\bibfnamefont{A.~K.} \bibnamefont{Hartmann}} \bibnamefont{and}
  \bibinfo{author}{\bibfnamefont{H.}~\bibnamefont{Rieger}},
  \emph{\bibinfo{title}{Optimization algorithms in physics}}
  (\bibinfo{publisher}{Wiley-VCH}, \bibinfo{year}{2002}).

\bibitem[{\citenamefont{Cherkassky and Goldberg}(1997)}]{Cherkassky-97}
\bibinfo{author}{\bibfnamefont{B.}~\bibnamefont{Cherkassky}} \bibnamefont{and}
  \bibinfo{author}{\bibfnamefont{A.~V.} \bibnamefont{Goldberg}},
  \bibinfo{journal}{Algorithmica} \textbf{\bibinfo{volume}{19}},
  \bibinfo{pages}{390} (\bibinfo{year}{1997}).

\bibitem[{\citenamefont{Hartmann}(1998)}]{Hartmann-98}
\bibinfo{author}{\bibfnamefont{A.~K.} \bibnamefont{Hartmann}},
  \bibinfo{journal}{PHYSICA A} \textbf{\bibinfo{volume}{248}},
  \bibinfo{pages}{1} (\bibinfo{year}{1998}).

\bibitem[{\citenamefont{Middleton}(2001)}]{Middleton-01}
\bibinfo{author}{\bibfnamefont{A.~A.} \bibnamefont{Middleton}},
  \bibinfo{journal}{Phys. Rev. Lett.} \textbf{\bibinfo{volume}{88}},
  \bibinfo{pages}{017202} (\bibinfo{year}{2001}).

\bibitem[{\citenamefont{Frontera et~al.}(2000)\citenamefont{Frontera,
  Goicoechea, Ort\'in, and Vives}}]{Vives-00}
\bibinfo{author}{\bibfnamefont{C.}~\bibnamefont{Frontera}},
  \bibinfo{author}{\bibfnamefont{J.}~\bibnamefont{Goicoechea}},
  \bibinfo{author}{\bibfnamefont{J.}~\bibnamefont{Ort\'in}}, \bibnamefont{and}
  \bibinfo{author}{\bibfnamefont{E.}~\bibnamefont{Vives}}, \bibinfo{journal}{J.
  Comp. Phys.} \textbf{\bibinfo{volume}{160}}, \bibinfo{pages}{117}
  (\bibinfo{year}{2000}).

\bibitem[{\citenamefont{Hartmann and Nowak}(1999)}]{Hartmann-99}
\bibinfo{author}{\bibfnamefont{A.~K.} \bibnamefont{Hartmann}} \bibnamefont{and}
  \bibinfo{author}{\bibfnamefont{U.}~\bibnamefont{Nowak}},
  \bibinfo{journal}{Eur. Phys. J. B} \textbf{\bibinfo{volume}{7}},
  \bibinfo{pages}{105} (\bibinfo{year}{1999}).

\end{thebibliography}

\end{document}